\documentclass[prl,aps]{revtex4}
\usepackage{latexsym}
\usepackage{amssymb}
\usepackage{amsmath}
\usepackage[varg]{txfonts}
\usepackage{mathrsfs}
\usepackage{upgreek}
\usepackage[]{subfig}
\usepackage{verbatim}
\usepackage{array}
\usepackage{color}
\usepackage{graphicx}
\DeclareMathAlphabet{\mathpzc}{OT1}{pzc}{m}{it}
\begin{document}
%\begin{frontmatter}
\title{Difference between stable and exotic nuclei: medium polarization effects}
%\author[1,2,3]{R. A. Broglia}
%\author[1,2]{G. Potel}
%\author[4]{F. Barranco}
%\author[2]{E. Vigezzi}
%\address[1]{Dipartimento di Fisica, Universit\`{a} di Milano,
%Via Celoria 16, 20133 Milano, Italy.}
%\address[2]{INFN, Sezione di Milano Via Celoria 16, 20133 Milano, Italy.}
%\address[3]{The Niels Bohr Institute, University of Copenhagen, Blegdamsvej 17,
%2100 Copenhagen {\O}, Denmark.}
%\address[4]{Departamento de Fisica Aplicada III, Universidad de Sevilla, Escuela Superior de Ingenieros,
%Sevilla, 41092 Camino de los Descubrimientos s/n,
%Spain.}
\author{R. A. Broglia}
\affiliation{Dipartimento di Fisica, Universit\`{a} di Milano,
Via Celoria 16, 20133 Milano, Italy.}
\affiliation{INFN, Sezione di Milano Via Celoria 16, 20133 Milano, Italy.}
\affiliation{The Niels Bohr Institute, University of Copenhagen, Blegdamsvej 17,
2100 Copenhagen {\O}, Denmark.}
\author{G. Potel}
\affiliation{Dipartimento di Fisica, Universit\`{a} di Milano,
Via Celoria 16, 20133 Milano, Italy.}
\affiliation{INFN, Sezione di Milano Via Celoria 16, 20133 Milano, Italy.}
\author{F. Barranco}
\affiliation{Departamento de Fisica Aplicada III, Universidad de Sevilla, Escuela Superior de Ingenieros,
Sevilla, 41092 Camino de los Descubrimientos s/n,
Spain.}
\author{E. Vigezzi}
\affiliation{INFN, Sezione di Milano Via Celoria 16, 20133 Milano, Italy.}

%\maketitle
\begin{abstract}
The $^1S_0$ phase shift is large and positive at low densities (relative momenta), while it vanishes and eventually becomes negative at densities of the order of the saturation nuclear density. The bare $NN$--potential, parametrized so as to reproduce these phase shifts leads to a sizable Cooper pair binding energy in nuclei along the stability valley. It is a much debated matter whether this value accounts for the ``empirical'' value of the pairing gap or whether a similarly important contribution arises from the exchange of collective vibrations between Cooper pair partners. In keeping with the fact that two--particle transfer reactions are the specific probe of pairing in nuclei, and that exotic halo nuclei like $^{11}$Li are extremely polarizable (representing, as far as this property is concerned, almost a caricature of stable nuclei), we find that the recent studied reaction, namely $^{11}$Li+$p\to ^9$Li+$t$, provides direct evidence of phonon mediated pairing in nuclei.
%, but it is not able to bind the single Cooper pair (halo neutrons) of $^{11}$Li
\end{abstract}
%\end{frontmatter}
\maketitle
\section{Introduction}
Arguably, one of the greatest achievements of many--body physics has been that of providing a complete description and a thorough understanding of superconductivity. At the basis of it one finds BCS theory and the Josephson effect. The first recognized the central role played by the appearance of a macroscopic coherent field usually viewed as a condensate of strongly overlapping Cooper pairs, the quasiparticle vacuum. The second made it clear that a true gap is not essential for such a state of matter to exist, but rather a finite expectation value of the pair field. Consequently, the specific experiments to study the superconducting state is Cooper pair tunneling. Such experiments allowed for a detailed probing of the phonon mediated pairing interaction leading to a thoroughly quantitative ``exact'' era in the study of pairing in metals, with uncertainties well below the 10\% level.

From this vantage point of view, it is not difficult to argue that major progress in the understanding of pairing in atomic nuclei --a subject lying at the forefront of nuclear research but still far away from having entered the ``exact'' era-- will arise from a systematic study of two--particle transfer reactions on drip line, exotic nuclei like e.g. $^{11}$Li, stabilized by the pairing correlations associated with a single Cooper pair.

 While in the infinite system the existence of a bound state of the Cooper pair happens for an arbitrarily weak attractive interaction, in the nuclear case this phenomenon takes place only if the strength of the interaction is larger than a critical value connected with the discreteness of the nuclear spectrum around the Fermi energy. In fact, in the case of $^{11}$Li the pairing interaction arising from the bare nucleon--nucleon potential seems not able to bind the halo neutrons, and one can posit that the exchange of collective vibrations between the Cooper pair partners are the main source of pairing in the low density systems (\cite{Barranco:01}). At the basis of this result is the low moment (large neutron mean square radius) and associate small value of $S_{2n}$ and thus: a) the low angular momentum content (essentially $s$--, $p$-- and $d$--waves) of the phase space available to the halo neutrons to correlate, and b) the high polarizability of the $^9$Li core and of the halo field. In such low--angular momentum phase space the two neutrons are not able to fully profit from the strong force--pairing, known to receive important contributions from many high--$l$ partial waves (\cite{Belyaev:59}, \cite{Mottelson:58}, \cite{Mottelson:62}). On the other hand, the collective $L=0,1$ and 2 fluctuations of the medium strongly renormalize the neutron motion leading to a bound Cooper pair (\cite{Barranco:01}).

\section{Minimal mean field approximation}
Classically, particles have definite positions, waves well defined momenta. Because of the complementary principle, the results of momentum and position measurements must fulfill the relation $\Delta x \Delta p_x \geq i\hbar$, tantamount of saying that particles can be described equally well as waves and viceversa (\cite{Bohr:28}), and that nothing is gained by discussing basic problems in terms of one rather than the other picture (\cite{Heisenberg:30}). One of the most important consequences of this shift of paradigm in the description of the physical world is the fact that, contrary to the classical picture where particles in empty space (vacuum) have well defined masses and charges, regardless of the rest of the universe, in quantum mechanics all physical properties can be viewed as collective properties (\cite{Laughlin:06}) or, more accurately, many--body properties. Quoting from Laughlin, ``A nice example of a collective effect \ldots is emitted light from dilute atomic vapors with special wavelengths so insensitive to outside influences that they con be used to make clocks accurate to one part in one hundred trillion. But these wavelengths have a detectable shift at one part in ten million --ten trillion times larger than the timing errors of the clock-- which should not have been present in an ideal world containing nothing but the atom (\cite{Lamb:50}, \cite{Lamb:51}) \ldots calculations then revealed this shift to be an electrical effect of the vacuum of space \ldots The ostensible empty vacuum space, in other words, is not empty at all but full of ``stuff''$\,$'' (see also \cite{Broglia:02c}).

From this vantage point of view, it is particularly illuminating for the subject of the present article to quote what Ben Mottelson (\cite{Mottelson:62}) wrote at the beginning of the modern era of nuclear structure: ``\ldots, in a many--body system such as the nucleus every feature is in some sense a collective phenomenon --every property depends on the total organization of the system and reveals the (collective) contribution of all nucleons. Indeed the most striking and fundamental collective picture in all nuclear phenomena is the existence of an average field in which the nucleons move approximately independently.'' The impact of these statements become even stronger by remembering the fact that the Hartree--Fock vacuum corresponds to a system in which all fermion levels are occupied up to the Fermi energy, similar to the Dirac vacuum (negative energy solutions of Dirac equation). The ZPF (particle--hole like phonon excitations) associated with the collective vibrations displayed by nuclei, renormalize in an important way the single--particle motion, giving rise to effective masses (density of levels around $\epsilon_F$) as well as to finite lifetimes for levels removed from the Fermi energy, as required by the experimental data.

\begin{figure}
\centerline{\includegraphics*[width=.55\textwidth,angle=0]{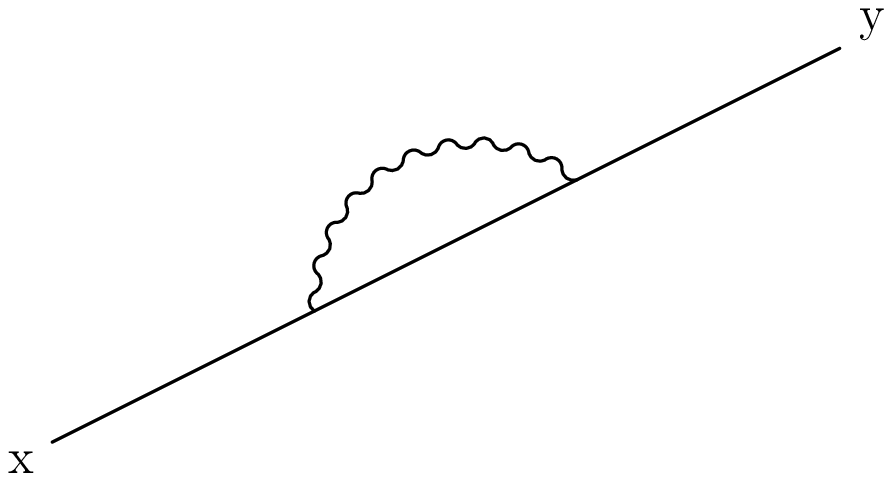}}
\caption{Lowest order self energy process of a particle with its own field.}\label{fig1}
\end{figure}

\begin{figure}
\centerline{\includegraphics*[width=.55\textwidth,angle=0]{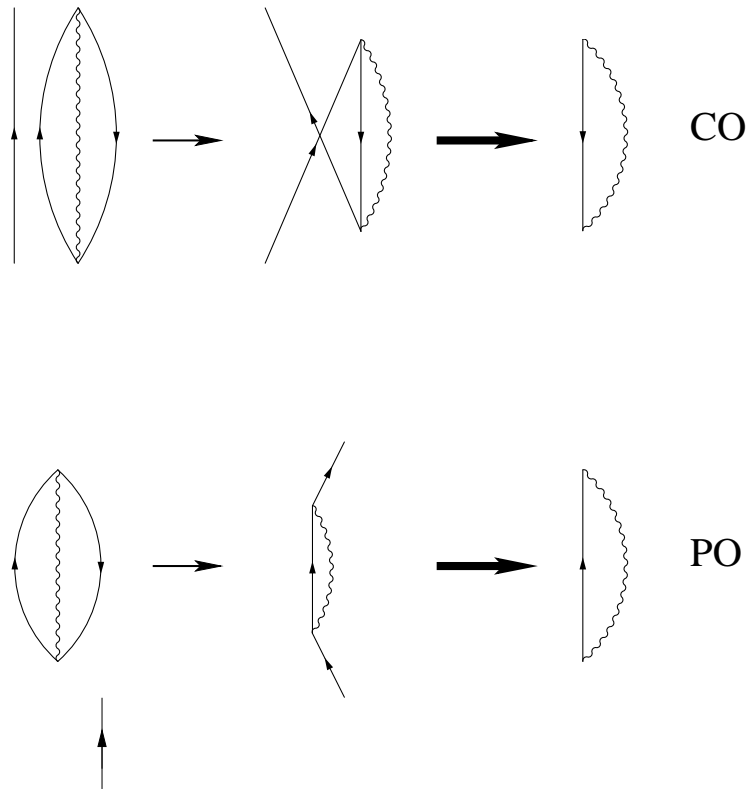}}
\caption{Relation between CO and PO processes and vacuum ZPF (oyster--like diagrams).}\label{fig2}
\end{figure}

This phenomenon is nothing else but another example that, as stated by Feynman (\cite{Feynman:75}) ``\ldots nothing is really free. For an electron going from $x$ to $y$ (see Fig. \ref{fig1}), the pole of the propagator for a free particle is at $p^2=m^2$. However, making measurements at $x$ and $y$ we could not tell if the electron had emitted and absorbed any number of photons (tantamount to say: being affected any number of times by the ZPF of the vacuum). Such processes, the simplest of which is shown in Fig. \ref{fig1}, cause a shift in the position of the pole. Physically, this means that what we measure (the ``experimental'' mass, $m_{exp}$) is not the ``bare'' mass, but something else which includes the effect of the virtual process mentioned above\ldots This discussion shows that the ``bare'' mass\ldots is in fact not directly observable.'' In other words, one can state that the different process which renormalize the single--particle motion, namely correlation (CO) and polarization (PO) processes arise from ZPF (see Fig. \ref{fig2}).

\begin{figure}
\centerline{\includegraphics*[width=.55\textwidth,angle=0]{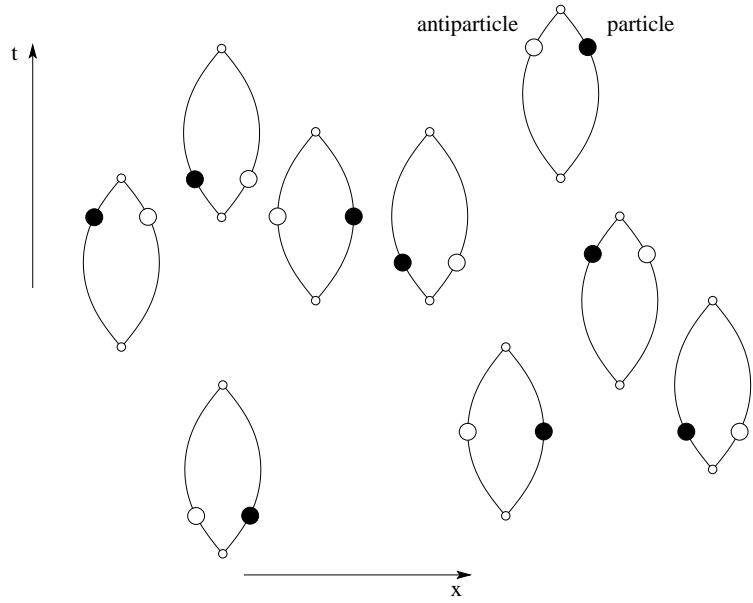}}
\caption{Schematic representation of quantal vacuum ZPF. Particle--antiparticle pairs are spontaneously (virtually) excited and annihilated.}\label{fig3}
\end{figure}

 Within this context one can mention that quantal ZPF are also having profound consequences not only at the level of the very small, like an atomic nucleus, but also at the level of the space time description of the Universe as embodied by the  theory of general relativity (\cite{Krauss:99}). At the heart of this theory one finds the field equation $G_{\mu\nu}=8\pi G T_{\mu\nu}$ which states that the geometry of spacetime, embodied in the curvature tensor $G_{\mu\nu}$, is determined by the distribution of matter and energy $T_{\mu\nu}$ (stress--energy tensor), where $G$ is Newton's constant. To create a static model of the universe, Einstein introduced in his equation a term proportional to the spacetime metric tensor $g_{\mu\nu}$ of strength $\Lambda$ (cosmological constant), to counterbalance gravity's attraction on cosmic scales. This cosmological term was added to the left side of the field equation, implying that it was a property of space itself. It was abandoned once it became clear that the universe was expanding. Recent work on quantum gravity has shown the need for a new cosmological term $\rho_{vac}$ (the energy density vacuum) associated with the quantal ZPF of the vacuum, again proportional to $g_{\mu\nu}$. This term is now to be added to the right hand term of the field equation, implying a form of energy which arises from virtual particle--antiparticle pairs (see Fig. \ref{fig3}).

In keeping with the above discussion, but now within the field of nuclear physics, one can posit that minimal mean field theory should contain, aside from the Hartree-- and the exchange (Fock--) potential, the (complex) self--energy contributions associated with ZPF renormalization processes (dynamical shell model, see \cite{Mahaux:85}).

%(see \ref{fig4}). In other words,
%\begin{equation}\label{eq1}
%    \begin{split}
%    \left[-\frac{\hbar^2}{2m}\nabla^2\right.&\left.+U(r)+\Delta E_j(\omega)\right] \varphi_j(\vec{r})\\
%    &+\int d^3r' U_x(\vec{r},\vec{r}')\varphi_j(\vec{r}')=\varepsilon_j\varphi_j(\vec{r})
%    \end{split}
%\end{equation}
%non--local in space and time
%\begin{equation}\label{eq2}
%U(r)\quad\text{local}\quad\text{Hartree}
%\end{equation}
%
%\begin{equation}\label{eq3}
%U_x(\vec{r},\vec{r}')\quad\text{nl space}\quad\text{Fock (exchange). Four. transf.} U_x(\vec{r},\vec{r}')\Rightarrow\tilde U_x(k)
%\end{equation}
%
%\begin{equation}\label{eq4}
%\Delta E_j(\omega)
%\end{equation}
%\emph{One way to make it local}: eff. mass and charge
%\begin{equation}\label{eq5}
%    \left[-\frac{\hbar^2}{2m^*}\nabla^2+U'(r)+\right] \varphi_j(\vec{r})=\varepsilon_j\varphi_j(\vec{r})
%\end{equation}
%\begin{equation}\label{eq6}
%    m^*=\frac{m_k m_\omega}{m} \quad(\approx m, \text{see next page})
%\end{equation}
%However, very different from Hartree where occupation of states is either 0 or 1. Here
%\begin{equation}\label{eq7}
%    Z_\omega=\left(\frac{m_\omega}{m}\right)^{-1}
%\end{equation}
%\begin{equation}\label{eq8}
%    m_\omega=m\left(1-\frac{\partial \Delta E(\omega)}{\partial(\hbar \omega)}\right)
%\end{equation}
%\begin{equation}\label{eq9}
%    m_k=m\left(1+\frac{m}{\hbar^2 k}\frac{\partial \tilde U_x(k)}{\partial k}\right)^{-1}.
%\end{equation}
In other words, in the simplest version of the nuclear shell model, it is assumed that the nucleons move independently of one another in a static mean field. This is necessarily an oversimplification of the physical reality, but many experiments indicate that it has some degree of validity. In fact, Hartree--Fock theory provides essentially the right sequence of single--particle levels and thus predicts, as a rule, the correct magic numbers for nuclei along the valley of stability. However, close to the Fermi energy (within a range of $\approx \pm 5$ MeV) it predicts a density of levels which is too low as compared with the experimental data. Furthermore, away from the Fermi energy ($\pm 10$ MeV), while the predicted HF density of levels is correct, it fails to account for the finite lifetime (width) experimentally observed. The above shortcomings underscore the need for a better theory. Not surprisingly, and using again an analogy from the physics of nature developed by Feynman (as well as Schwinger and Tomonaga (\cite{Schweber:94})) the whole of the physics of such a theory corresponds to take into account vacuum fluctuations as has been done in the in going from the standard (Schr\"{o}dinger, Dirac) description of the atom, to QED.

 In fact, in HF theory, the vacuum (ground state) corresponds to a determinant with all levels below the Fermi energy occupied, those above empty. Now, such a system can vibrate with particle--hole as well as particle--particle like collective modes which, as a rule behave like quasi harmonic modes. The associated ZPF correspond to virtual excitations, a process described, in lowest order, by an oyster type diagram (see Figs. \ref{fig2} and \ref{fig4}(c')). Such a process renormalizes the single--particle levels leading to the correct density around the Fermi energy, and a breaking of the single--particle state far away from the Fermi energy. In other words a minimal, self--consistent theory of independent particle motion is achieved when one considers the fact that in their motion, nucleons interact with the rest of the nucleus and drag core excitations, leading to dispersion corrections of the single--particle motion known as polarization and correlation effects (see Fig. \ref{fig2}). One then speaks of the motion of quasiparticles.

In other words, quasiparticles based on single--particle levels with energies close to the Fermi energy display a group velocity smaller than that of particles with the same wavelength. This effect is measured by both the $\omega$--mass $m_\omega(E)$ which is a function of the energy $E$ of the quasiparticle, and by $Z_\omega(E)=m/m_\omega(E)$, the discontinuity of the Fermi energy, closely connected with the spectroscopic factor as measured in single--particle transfer reactions. For quasiparticles associated with single--particle levels displaying energies far removed from the Fermi energy, medium polarization effects give rise to inhomogeneous damping. Such a phenomenon is characterized by the FWHM $\,\Gamma_\omega(E)$, which provides the range of energies over which, due to the particle--vibration coupling mechanism, the single--particle strength is distributed. Of notice that this phenomenon does not give rise to a proper lifetime of the quasiparticle state but to a fractionation of the single--particle strength.

Because the main contributions to medium polarization effects arise from the coupling of nucleons to low--lying surface collective vibrations ($\hbar\omega_\lambda\approx$ few MeV), one expects that $m_\omega(E)$ displays a well defined peak around $\epsilon_F$, in keeping with the fact that once $E-\epsilon_F$ is larger than $\hbar\omega_\lambda$, the collective excitations are decoupled from the particle, whose group velocity therefore increases. There is strong experimental evidence which testifies to this scenario. Now, in many--particle systems, virtual phonons can, not only be emitted and reabsorbed by the same fermion, but also be exchanged between two fermions, giving rise to an induced interaction (medium polarization effects).

 We know that such effects play a central role in the phenomenon of superconductivity in metals at low temperatures. In keeping with the analogy employed by \cite{Bohr:58} to justify the use of BCS theory to explain the energy gap observed in the intrinsic excitation spectrum of spheroidal nuclei, an analogy which started the field of nuclear superfluidity, it seems fair to ask how important medium polarization effects are in nuclear pairing. In fact, in keeping with the dynamical shell model discussed above, the question is not whether one has to consider the pairing interaction arising from the exchange of vibrations between nucleons moving in time reversal states close to the Fermi energy, but what the values of the associated matrix elements are as compared with those associated with the bare nucleon--nucleon interaction.
\section{A change in paradigm}
Although not explicitly stated, last section provides evidence for a change of paradigm in the treatment of physical systems, ranging from the Universe to the atomic nucleus. From one of a static, symmetry dominated scenario (geometry of space time, mean field of atoms and nuclei), to another in which the properties of a system do not depend only on the particles which form it, nor on the forces acting among them but also, and primarily, on the medium in which they are embedded (and thus on the $\omega$-- and $k$-- dependent generalized dielectric function). And by (external) medium it is meant both that felt by the two weakly bound neutrons of $^{11}$Li, namely the halo field created by themselves, as well as the vacuum (field) felt by an atom or by the whole universe, field which its own expansion creates.
 To these fields are associated ZPF. It is similar to a quantal harmonic oscillator, in which each degree of freedom contributes a quantity $(1/2)\hbar\omega_0$ to the ground state energy, the associated ZPF being measured by $(\hbar \omega/2C)^{1/2}$.

 In the case of $^{11}$Li, these quanta are essentially the dipole (pigmy) resonance and the quadrupole mode. In the case of the electromagnetic vacuum, the photon. More generally, on the very small scales where quantum effects become important, even empty space is not really empty. Instead virtual particle--antiparticle pairs pop up of the vacuum, travel for short distances and then disappear again on timescales such that one cannot observe them directly (see Fig. \ref{fig3}). Yet their indirect effects are very important, and can be measured. For example, in the Lamb shift, as well as in the Casimir effect to name but two.

\begin{figure}
\centerline{\includegraphics*[width=.75\textwidth,angle=0]{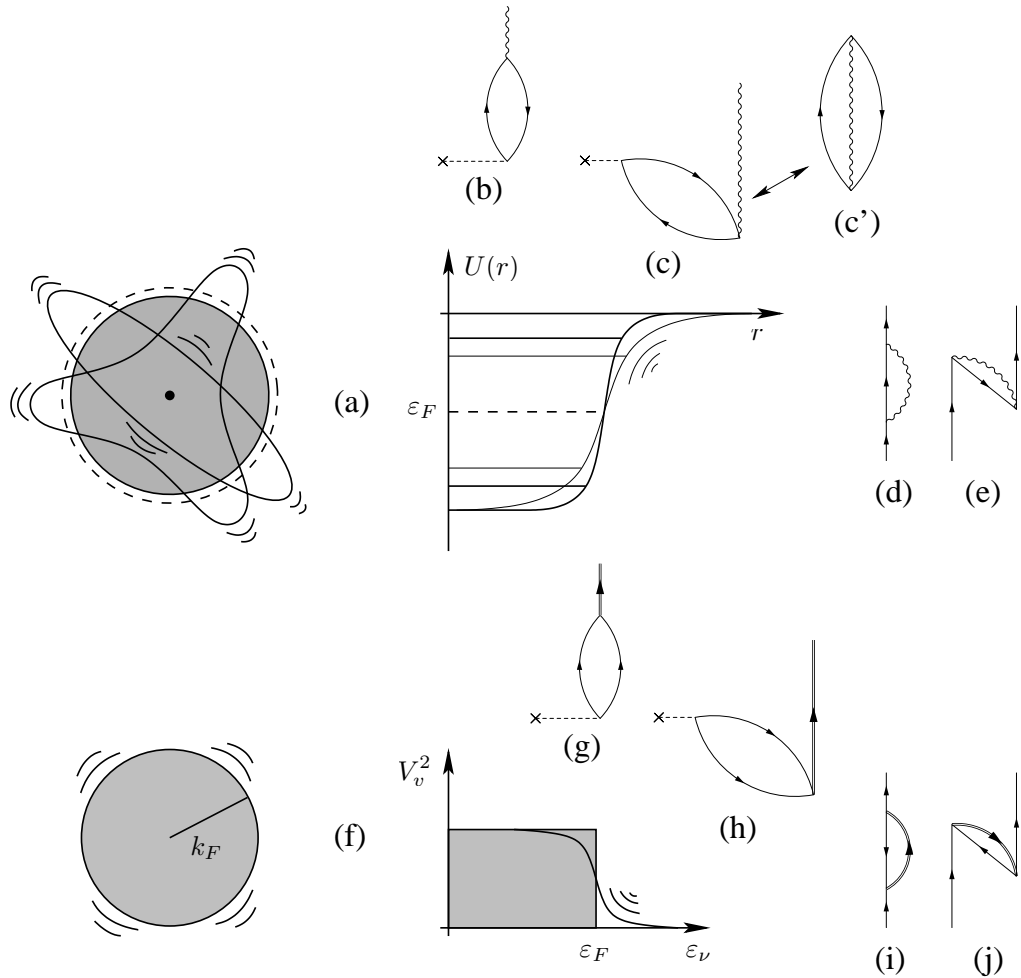}}
\caption{Schematic representation of the surface (a) and of pairing (doubled arrowed line) (f) vibrations.}\label{fig4}
\end{figure}

Returning to the nuclear case, mean field defines, among other things, a surface. This surface can vibrate collectively, the instantaneous distortion of it, associated with the ZPF of the collective modes, point at each instant of time, into a different direction, thus averaging out the dynamic braking of symmetry they introduce. Within this scenario, the ground state mean field properties (mean square radius, binding energy, density of levels  at the Fermi energy, etc.) of the system under study will be renormalized by the ZPF. In fact, a finite system which fluctuates (both radius and diffusivity) will, effectively display varied properties from those associated with the original static, mean field solution (\cite{Esbensen:83}, \cite{Barranco:85}, and Figs. \ref{fig4}(a)--(e)). Of notice that the same picture can be used to describe rotations, by keeping inertia finite and making the restoring force goes to zero. In this case, the ZPF correspond to an averaging of the angles between the laboratory reference frame, and the privileged orientation defined by the nuclear deformation. Because these angles vary over the whole range of possible values (in keeping with the fact that the restoring force associated with orientation is zero), is that in this case the ZPF diverge. The associated rotational mode is the equivalent, in the case of finite nuclei, of the Anderson--Goldstone--Nambu mode of field theories displaying spontaneous symmetry breaking. Thus, it is intimately connected with symmetry restoration. Neglecting ZPF implies, in this case, violation of angular momentum invariance, a possibility not contemplated by quantum mechanics. In keeping with this fact one can posit that the same is true in the case of finite (as opposite to divergent) values of the ZPF (vibrations with finite restoring force). In fact, no new physics is introduced in going from finite to vanishing restoring force. If one cannot neglect ZPF in deformed systems (angular momentum projection), one cannot do it either in the case of finite amplitude vibrations. The same arguments apply as well to pairing vibrations and rotations, that is, collective modes associated with spontaneous symmetry breaking of particle number.

The same can be argued concerning pairing modes. In fact, mean field not only determines a surface in 3D--space, but also in $k$--space (Fermi surface), which defines a number of particles (Fermi energy). Fluctuations in it lead to pair addition and to pair substraction modes (\cite{Bes:66} and Figs. \ref{fig4}(f)--(j)) and eventually pairing rotations when the associated restoring force for pairing vibrational modes vanishes.
 The coupling of pairing vibrations to single--particle motion can affect alignment in deformed rotating nuclei (see Fig. \ref{fig4}(i)) as well as the value of the pairing gap in an important way (see \cite{Shimizu:89}, \cite{Barranco:87b}).

\begin{figure}
\centerline{\includegraphics*[width=.55\textwidth,angle=0]{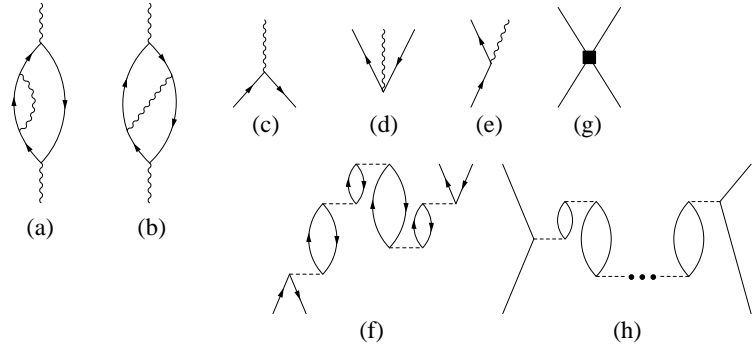}}
\caption{Self--energy processes of particle hole vibration ((a) and (b)) RPA ((c), (d)) and scattering (e) particle--vibration coupling vertices. In (f) we display the RPA diagrams in terms of $p$--$h$ bubbles. (g) Four point vertex, (h) particle--particle (pairing) interaction mediated by collective $p$--$h$ vibrations.}\label{fig5}
\end{figure}

 Of notice that when one calculates, on a mean field basis, the nuclear linear response, that is, collective $p$--$h$ like modes, the only consistent way to avoid talking about ZPF, is to use the Tamm--Dancoff approximation (see e.g. \cite{Rowe:70} and Fig. \ref{fig4}(b)). Now, we know that such approximation is not physically correct, as it violates the energy weighted sum rule (EWSR), which, in the case of the (electric) dipole mode is tantamount to saying violation of (charged) particles (Thomas--Reiche--Kuhn sum rule). This is the main reason why linear response is calculated in the RPA (or QRPA for superfluid nuclei) taking into account ground state correlations. As seen from Fig. \ref{fig4}(c) this is equivalent at saying that one can excite equally well collective modes through a direct promotion of a particle from a level lying below to Fermi energy (Fig. \ref{fig4} (b)) to one above then by forcing, with the help of an external field, the virtual ZPF (oyster diagram, see Fig. \ref{fig4}(c')) from being virtual to become real. Now, as seen from Figs. \ref{fig4}(d) and (e) (see also Fig. \ref{fig2}), a nucleon in presence of ZPF becomes dressed, that is becomes a real nucleon, whose properties can be confronted with experiment (effective--mass ($\omega$--mass $m_\omega$), --occupation ($Z_\omega=(m/m_\omega)$), --charge, etc.). In other words, the same process which is needed to fulfill the EWSR (Figs. \ref{fig4}(b),(c)), implies that nucleons should be dressed (Figs. \ref{fig4} and \ref{fig4}(d) and (e)). Consequently the RPA use in the calculation of collective modes is intrinsically wrong, as the backwards going amplitude $Y^{\lambda}_{ph}$ requires the consideration of ZPF which inescapably leads to effective mass process not considered in RPA (nor in QRPA). Consequently, the minimal description of collective modes is that shown with the diagram of Fig. \ref{fig5}(a). Now, this diagram is one of the many which describes the coupling of one-- with two--phonon states. Because of Furry's theorem, it necessarily requires that vertex correction processes (see Fig. \ref{fig5}(b)) be taken into account on equal footing than self--energy ones. More appropriate within the framework of Nuclear Field Theory (NFT) (\cite{Bes:76c}, \cite{Bortignon:77}, \cite{Mottelson:76}) context, the diagrams shown in Figs. \ref{fig5}(a) and \ref{fig5}(b) are both necessary to satisfy generalized Ward identities (see e.g. \cite{Schrieffer:64}).

\begin{figure}
\centerline{\includegraphics*[width=.55\textwidth,angle=0]{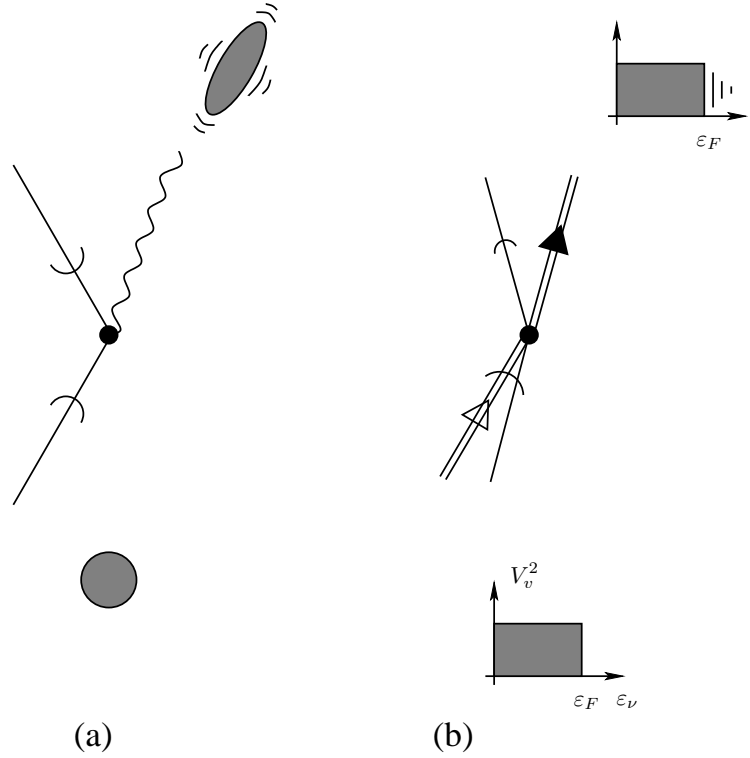}}
\caption{Diagramatic representation of: a) inelastic scattering process and b) two--particle transfer process.}\label{fig6}
\end{figure}

Stating the same concept but in an even simpler way one can posit that of all the three possible particle--vibration coupling vertices (see Figs. \ref{fig5}(c), (d) and (e)) RPA (or QRPA) selects only two, namely \ref{fig5}(c) and \ref{fig5}(d). This is because RPA is a harmonic approximation. In other words, one assumes the collective vibration to be a phonon state of an harmonic oscillator Hamiltonian. Thus the coupling between one-- and two--phonon states must be zero. This is guaranteed if one does not consider the scattering term shown in Fig. \ref{fig5}(e). Now, this is a contradiction in terms, in keeping with the fact that it is through an inelastic scattering experiment, as displayed in Fig. \ref{fig6}(a) that one can measure the particle--vibration coupling vertex, or better the transition density of the RPA mode (same for pairing vibrations, see Fig. \ref{fig6} (b)). But if one has eliminated graph \ref{fig5}(e) in the calculation of the mode (of notice that single arrowed lines pointing upwards represent bound particles), one can hardly imagine that the vertex becomes operative only because the nucleon is now in a scattering state (single (curved) arrowed line pointing upwards, Fig. \ref{fig6}(a)).

Summing up, it is certainly very important to try to use the best available four--point vertex (including also three--body interactions) in the calculation of the process displayed in Fig. \ref{fig5}(f), as well as those associated with the pairing interaction (Fig. \ref{fig5} (g) and (h)) taking properly into account, aside from the standard central term, the isovector, the spin--spin, tensor, etc. components. However, without medium polarization effects of the type displayed in Figs. \ref{fig5}(a) and (b) as well as (h), the corresponding theoretical description is not consistent, as e.g. QED is in the description of electromagnetic processes. Even worst, one may be able to fit some data. However sooner or later the need to include fluctuations on par with mean field properties will be forced by experiment. As we shall see below in connection with the (pairing) particle--particle channel (see Fig. \ref{fig5}), by the direct observation of individual quanta of the glue acting among Cooper pairs. Surprisingly, while any modern calculation of the pairing gap in neutron and nuclear matter considers not only the bare $NN$--interaction (or whatever effective force which its place), but also the induced pairing interaction, a clear resistance to consider medium fluctuation effects in the case of pairing in finite nuclei is apparent.

 On the other hand, this resistance of today practitioners has in some sense a touch of d\'{e}j\`{a} vu. In fact the surprise and, to some extent resistance caused by the advent of the nuclear shell model (Goeppert Meyer--Jensen) which, apparently, so directly contradicted the liquid drop and compound nucleus model developed by Niels Bohr and collaborators, triggered a number of (static mean field like) theoretical explanations mostly based on the Pauli principle. The limitations of such an approach has been forcefully argued by Ben Mottelson (\cite{Mottelson:02}) making use of the quantality parameter $Q=(\hbar/Ma^2)/|v_0|$ which provides a measure of the validity of independent particle motion ($Q\ll 1$ implies localization, while $Q\gtrsim 1$ is tantamount to delocalization, an example of the fact that, while potentials prefer definite relations among particles, fluctuations, quantal or classical, favour symmetries). Typical values of the parameter defining the bare $NN$--potential ($a=1$ fm, $v_0=-100$ MeV) lead to $Q\approx 0.4$. That associated with the induced interaction medium polarization effects arising from the exchange of collective vibrations between nucleons moving close to the Fermi energy ($a\approx 10$ fm, $v_0\approx -1$ MeV) are associated with  $Q\approx 4$. While the value $Q\approx 0.4$ can be assigned to stable nuclei, the second, largest value is more representative of halo nuclei like $^{11}$Li and  $^{12}$Be. Consequently, one expects exotic, halo, nuclei to provide an excellent testing ground to study the role virtual processes play in the renormalization of single--particle motion and of $NN$--interactions, in particular pairing $(^1S_0)$ interaction (see Fig. \ref{fig5} (h)). One-- and two--particle transfer reactions which make real such virtual processes, can be used as the specific probes to learn about the highly polarizable fragile objects known as halo nuclei, thus shedding light on what, arguably, can be considered the essence of finite, quantum many--body systems: zero point fluctuations.

  Let us conclude this section by reminding that there is an essential difference between the concepts of symmetries and that of spontaneous symmetry breaking (and of its intrinsically connected modes leading to symmetry restoration). The first one provides a static, geometric view of the phenomenon. The second one not only considers paramount all and each one of the symmetries of the Hamiltonian describing the system, but at the same time and on equal footing the fluctuations of the mean field responsible for the dynamic maintenance of the associated invariance (or symmetry restoration). In other words, spontaneous symmetry breaking  not only takes fluctuations into account but is, as a rule, associated with a boson (Anderson--Goldstone--Nambu mode) for which the fluctuations are not only large, but diverge.

Of notice that the static $\to$ dynamic change of paradigm in the study of complex systems is being observed in other fields of research different than physics, like e.g. in the study of proteins and protein inhibitors (see e.g. \cite{Broglia:08}). Proteins, the hardware of life (metabolism and structure), are linear chains of amino acids produced by the ribosome. They fold in short times (typically milliseconds) to their native, biologically active structure. In particular enzymes which, in the native conformation display an active (catalysis inducing) site. In other words, proteins to work have to be folded. The static, symmetry driven picture of nature, see proteins as folded proteins. Not surprisingly, conventional drugs are designed to bind to the active site thus blocking activity, and consequently the ability of the patogen agent (virus, bacteria, etc.) for which the target enzyme plays a central role, to mature and eventually reproduce. Mutations, as a rule, distorting the active site, create resistance. In certain cases (e.g. in connection with inhibitors of the viral hepatitis C) within days from the beginning of therapy.

Studying the mechanism through which a protein, starting from its denatured state folds into the native conformations, one can develop inhibitors which block folding, that is the dynamic process, by binding to those segments of amino acids which, playing a central role in the folding process, cannot be mutated. As a consequence, folding inhibition is likely not to create resistance, opening new venues to deal with infectious diseases. In other words, also in the field of biology and biochemistry and, as a result of it, in the design of drugs, a shift of paradigm is taking place, recognizing that proteins are dynamic systems folding as well as displaying conspicuous fluctuations, also in the native (ground state).

\section{Medium polarization pairing interaction in nuclei}
There are a couple of facts which have made the discussion of bare versus induced pairing interactions in nuclei difficult:\\ 1)The bare $NN$--interaction leads, as a rule, to a solution of the BCS (or to the HFB) equations with a finite value of the pairing gap (at variance with the case of metal superconductivity, where the (screened) Coulomb repulsion among electrons leads to values of $\Delta=0$).\\
2)Although the pairing gap is an important quantity in the discussion of superfluidity and superconductivity, it is not the specific observable of these phenomena. In fact, the specific probe of pairing correlations in nuclei is two--particle transfer reactions, the associated absolute cross section being indirectly related to the pairing gap.
Not surprisingly, we have found (\cite{Potel:09b}) that a recent two--particle transfer experiment provides direct, quantitative evidence of the role played in pairing correlations by the exchange of collective modes between Cooper pair partners as had been predicted few years ago (see \cite{Brink:05}, Ch. 11).

A multipole expansion of the $NN$--pairing interaction reveals that none of the multipole terms is more important the other, and that high multipoles have to be considered to achieve convergence. One can thus quench the importance of this interaction by concentrating on a nucleus in which the phase space allowed to them to correlate have small angular momentum content. A second condition is that of choosing a system which is very polarizable. It is argued in the next section that such a system are light halo nuclei in general and $^{11}$Li in particular.

\section{Exotic versus stable nuclei, direct observation of phonon mediated pairing}
\begin{figure}
\centerline{\includegraphics*[width=.55\textwidth,angle=0]{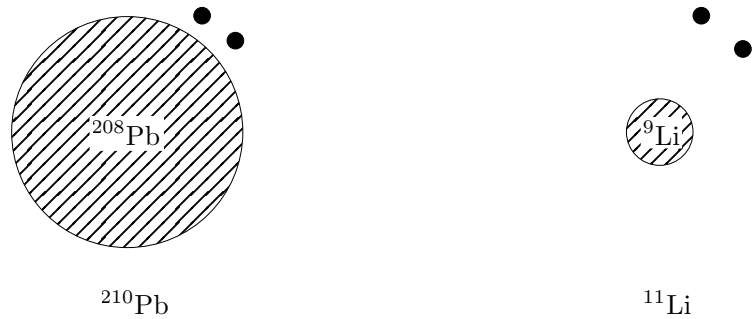}}
\caption{Schematic representation of the stable $^{210}$Pb and of the exotic (halo) nucleus $^{11}$Li viewed as a system with two neutrons (solid dots) moving around a core (dashed area). The distance of the valence neutrons from the core reflects the fact that the two neutron separation energy of these nuclei is $S_{2n}=15.2$ MeV and 0.380 MeV respectively.}\label{fig7}
\end{figure}

In \cite{Kuo:97} et al it is stated that exotic nuclei, being much less bound than stable nuclei, offer a unique framework to study mean field properties  without the complications of medium polarization effects. Paraphrasing the paper's arguments with the help of a cartoon representation of $^{210}$Pb and of $^{11}$Li (Fig. \ref{fig7}), one could argue that valence neutrons in Pb can exert a stronger polarization of the core than in Li, because they are closer to it.

 Now, nucleons in a nucleus resents not only of the bare $NN$--force, but also of medium polarization effects. In other words, the $NN$--force is modified by the nuclear dielectric function. Microscopically, this means that nucleons in a nucleus not only exchange pions but also nuclear vibrations. The fact that the full Thomas--Reiche--Kuhn (dipole) sum rule is concentrated in Pb at about 14 MeV, while as much as 20\% of it is found around 1 MeV in $^{11}$Li (pigmy resonance), constitutes a sobering warning concerning which of the two systems is more or less polarizable. In fact, microscopic calculations suggest that while the relative contribution to the pairing interaction associated with the bare $NN$--potential and with the exchange of collective vibrations is fifty--fifty in the case of stable nuclei (\cite{Barranco:99}, \cite{Barranco:04}), it is more 20\%--80\% in the case of light exotic, halo nuclei ($^{11}$Li, $^{12}$Be) (\cite{Barranco:01}, \cite{Gori:04}). At this point it is natural to ask about the experimental evidence which, specifically, can test these predictions.

\begin{figure}
\centerline{\includegraphics*[width=.8\textwidth,angle=0]{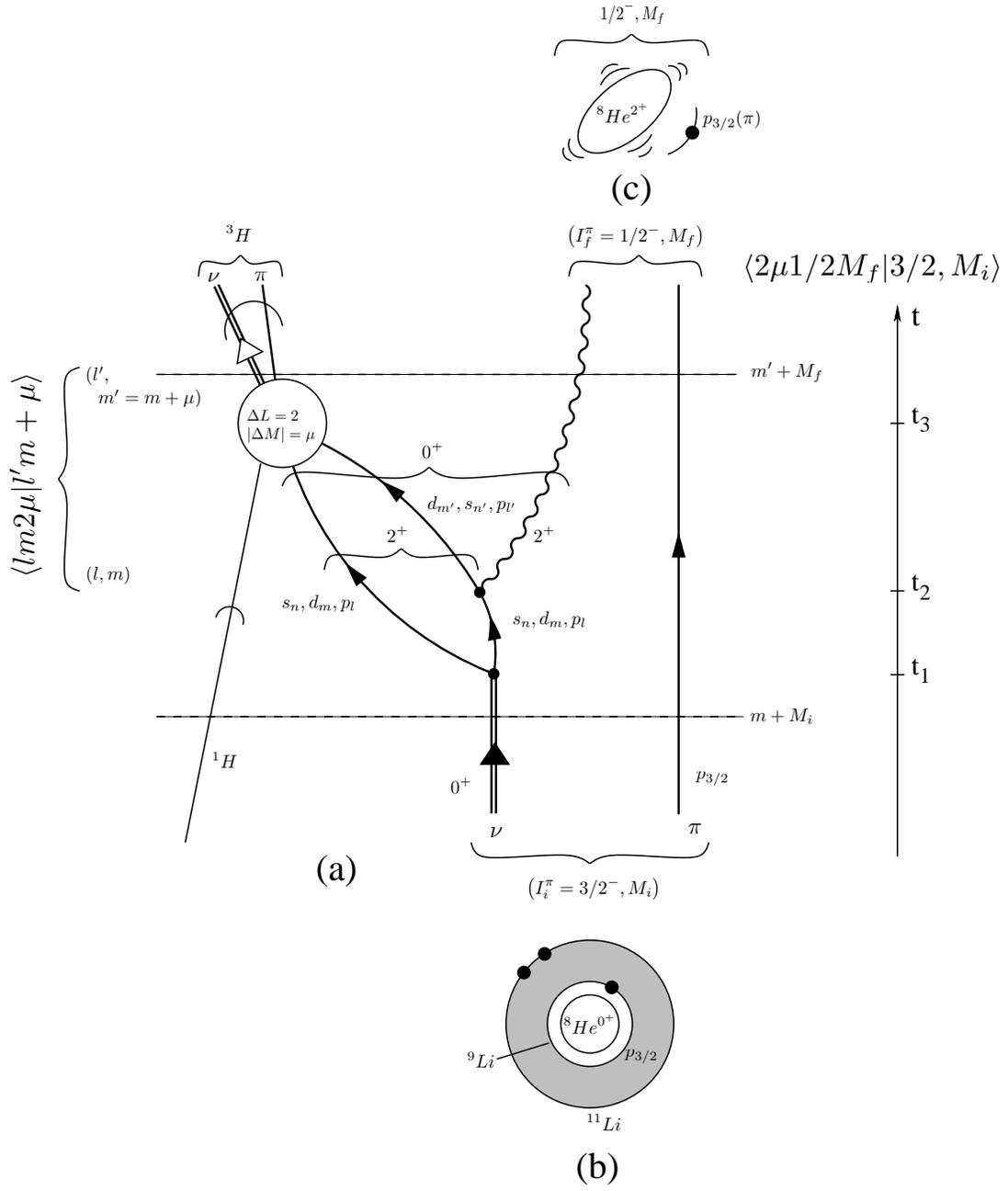}}
\caption{(a) NFT--Feynman diagram associated with the process \mbox{$^1$H ($^{11}$Li(gs),$^9$Li($1/2^-$;2.69 MeV)$^3$H)}, which treats on equal footing the nuclear structure
 and the reaction mechanism.
  Arrowed lines indicate bound particles, curved arrowed lines, scattering states.
Curly brackets indicate angular momentum coupling, while horizontal dashed lines indicate magnetic quantum number conservation
In (b) and (c) a schematic representation of the initial ($^{11}$Li) and final ($^9$Li($1/2^-,M_f$; 2.69 MeV)) nuclear states are given, respectively.}\label{fig8}
\end{figure}

  Recently the reaction  $^{11}$Li($p,t$)$^9$Li has been studied (\cite{Tanihata:08}). Aside from the ground state, the first excited state of $^9$Li($1/2^-$, $E_x=2.69$ MeV) was populated. The 2.69 MeV state of $^9$Li is the $1/2^-$ member of the multiplet ($2^+ \otimes p_{3/2}(\pi)$), the $|2^+\rangle$ being the quadrupole, particle--hole like vibration of $^8$He.  Assuming a direct reaction process, it can be seen from Fig. \ref{fig8}, that the reaction $^{11}$Li$+p\to^9$Li($1/2^-)+t$ provides direct evidence of the exchange of quadrupole phonons between the halo neutrons (\cite{Potel:09b}). The absolute value of the associated cross section is consistent with an amplitude of 0.1 of the $|\left((s,p)_{2^+}\otimes 2^+\right)_{0^+}p_{3/2}(\pi);3/2^-\rangle$ configuration in the ground state wavefunction $^{11}$Li, in overall agreement with the prediction of ref. \cite{Barranco:01}.

   In the last sentence of their paper, \cite{Tanihata:08} state two things, namely: 1) that the population of the first excited state of $^9$Li suggests a $1^+$ or $2^+$ configuration of the halo neutrons; 2) that this shows that a two-nucleon
transfer reaction as they study may give new insight in the halo structure of $^{11}$Li. We cannot
emphasize strongly enough our support  for such  statements.
Concerning the first one, because NFT studies of the structure of $^{11}$Li (\cite{Barranco:01}) indicated
this to be the only mechanism for $^{11}$Li to be bound, at the risk of questioning all what
has been learned concerning the workings of the particle-vibration coupling mechanism (see e.g. \cite{Bohr:75}, \cite{Mahaux:85}, \cite{Bes:76c}, \cite{Mottelson:76}, \cite{Bortignon:77}).
In reference to the second one, because it is high time to rediscover that  pair transfer is the specific probe of
pairing in nuclei (\cite{Bohr:75}, \cite{Broglia:73}).
\section{Conclusions}
The essence of finite many--body systems, like e.g. the atomic nucleus, are fluctuations. The interweaving of single--particle motion and the collective vibrations is required by sum rule arguments. Less bound a system is, more important polarization effects are expected to be. It is then not surprising that by studying a very fragile object like $^{11}$Li through a two--particle transfer experiment, specific probe of pairing, one has obtained the first direct evidence of phonon mediated Cooper pair binding in nuclei. Such a result indicates that while medium polarization effects pervasively renormalize all of the observables, they may be directly observed only in specific experiments. On the other hand, their effect, virtually, is present at all times. The sooner one comes to terms with this fact the better.\\ \\\emph{The authors want to acknowledge important discussions with Thomas Duguet concerning the pairing $NN$--interaction. His work and that of his collaborators on the subject are likely to be instrumental in ushering the field of nuclear superfluidity into its really quantitative era.}

%\bibliography{C:/Gregory/Broglia/notas_ricardo/nuclear_bib}

\newpage
%\begin{figure}
%\centerline{\includegraphics*[width=.8\textwidth,angle=0]{fig_8.ps}}
%\caption{(a) NFT--Feynman diagram associated with the process \mbox{$^1$H ($^{11}$Li(gs),$^9$Li($1/2^-$;2.69 MeV)$^3$H)}, which treats on equal footing the nuclear structure 
%($\corfermion^{2^+},\boson^{2+},\fermion$)
% and the reaction mechanism
% ($\contfermion, \conttritium$).
%  Arrowed lines indicate bound particles, curved arrowed lines, scattering states.
%Curly brackets indicate angular momentum coupling, while horizontal dashed lines indicate magnetic quantum number conservation
%In (b) and (c) a schematic representation of the initial ($^{11}$Li) and final ($^9$Li($1/2^-,M_f$; 2.69 MeV)) nuclear states are given, respectively.}\label{fig8}
%\end{figure}

\end{document}